# Pressure dependent electronic properties of MgO polymorphs: A first-principles study of Compton profiles and autocorrelation functions

K.B. Joshi[†], B.K. Sharma[‡], U. Paliwal[†] and B. Barbiellini[$]

[†]*Department of Physics, M.L. Sukhadia University, Udaipur-313001 (India).*

[‡]*Department of Physics, University of Rajasthan, Jaipur-302004 (India).*

[$]*Department of Physics, Northeastern University, 110 Forsyth St., Boston, MA 02115 (USA).*

[†]Corresponding author: E-mail; k_joshi@yahoo.com,

Phone: 91-294-2423641, Fax: 91-294-2423639

## Abstract

The first-principles periodic linear combination of atomic orbitals method within the framework of density functional theory implemented in the CRYSTAL06 code has been applied to explore effect of pressure on the Compton profiles and autocorrelation functions of MgO. Calculations are performed for the B1, B2, B3, B4, B8$_1$ and *h*-MgO polymorphs of MgO to compute lattice constants and bulk moduli. The isothermal enthalpy calculations predict that B4→B8$_1$, *h*-MgO→B8$_1$, B3→B2, B4→B2 and *h*-MgO→B2 transitions take place at 2, 9, 37, 42 and 64 GPa respectively. The high pressure transitions B8$_1$→B2 and B1→B2 are found to occur at 340 and 410 GPa respectively. The pressure dependent changes are observed largely in the valence electrons Compton profiles whereas core profiles are almost independent of the pressure in all MgO polymorphs. Increase in pressure results in broadening of the valence Compton profiles. The principal maxima in the second derivative of Compton profiles shifts towards high momentum side in all structures. Reorganization of momentum density in the B1→B2 structural phase transition is seen in the first and second derivatives before and after the transition pressure. Features of the autocorrelation functions shift towards lower r side with increment in pressure.

# 1. Introduction

Magnesium oxide is one of the representative materials, which has been found to play an important role in the geophysical models of the mantle-core boundary [1-3]. This stable binary oxide is structurally rich and crystallises in a number of polymorphs [4-14]. There is a resurgence of interest in the structural studies of MgO due to contradictory results on the thermal and pressurewise stability. The ambient pressure melting point is not precisely determined and have diverging opinions about the phase from which it transforms[15].The volume discontinuity at 170 GPa in the shock wave experiments remains unresolved [16]. In the quest of this, polymorphs of MgO, other than B1 and B2, are being explored. However, the focus has been to report structural and electronic properties of these polymorphs using atomistic simulations. Phase diagram and pressure induced structural phase transitions among various polymorphs are also studied from first-principles methods [10-16]. Some of the efforts are directed to study transitions both experimentally as well as theoretically. Remaining are studied either experimentally or theoretically.

Electron momentum density (EMD) is an important ground state property of materials. The quantity is directly related to the solid state wave functions in the ground state [17-23]. This can be accessed through 2D-ACAR experiments which offer good resolution [20-23]. Using position-sensitive detectors one can avoid the integration over planes in momentum space [20-23]. However, one needs to take care of electron-positron interactions while interpreting experimental results. Moreover, sensitivity of the positrons to defects in the solid causes limitations. Furthermore, in some materials role of positrons is crucial due to the Coulomb repulsion by the nuclei [20,21]. Compton scattering on the other hand does not suffer from these drawbacks [17-23]. Within the validity of impulse approximation, the technique has been applied to determine projection of EMD along the direction of the scattering vector. The $\gamma$-ray Compton scattering experiments offer limited resolution. With the availability of synchrotron radiation sources it is possible to perform Compton profile measurements at high resolution with very good statistics [19]. Moreover, availability of polarized photons using multiple wigglers enables to study spin dependent momentum density distribution in magnetic materials [19,24-26]. Change in the spin dependent momentum density applying temperature has also been explored [26]. Thus Compton scattering has been applied to unfold Fermi surface topology, nature of bonding, directional bonding behavior in a variety of solids [17-19,24-26]. Magnetic Compton scattering has been applied to study field and temperature dependent magnetic properties, distribution of magnetic moment, separation of orbital and spin part of the magnetic moment in novel rare earth materials [24-26]. Despite such a success there are rare attempts to study pressure dependent EMD or the Compton profiles. In the experiments, pressure exerting assemblies such as diamond anvil cell are necessary. The pressure assembly poses serious problems in background measurements. Nevertheless, a few experiments are performed to see effect of pressure on Compton profiles [27-30]. Thereafter, there is a scarcity of attempts to study effects of pressure on EMD, Compton profile or the autocorrelation functions (AF) experimentally or theoretically. This is probably the first attempt to report pressure dependent Compton profiles, their first and second derivatives and the autocorrelation functions from first principles method.

In this article, our major goal is to see effect of pressure on the Compton profiles and autocorrelation functions of the ionic solid magnesium oxide. It crystallises in a number of structures and show pressure induced structural phase transitions theoretically. Moreover, MgO has been studied extensively from the Hartree-Fock (HF) linear combination of atomic orbitals (LCAO) method using various versions of the CRYSTAL code [5,6]. These studies have well set the stability of the basis sets of magnesium and oxygen for the current investigations. The directional and isotropic Compton profiles of MgO in rocksalt i.e. B1 structure which is stable structure under ambient conditions are studied by a number of workers [31-34]. However, in the current effort we apply the LCAO method within the framework of density functional theory embodied in the CRYSTAL06 code [35-38]. In this endeavor, first we compute the lattice constants and bulk moduli of MgO crystallizing in the B1, B2, B3, B4, B8$_1$, and *h*-MgO structures by coupling the total energy calculations with the Murnaghan equation of state. These calculations are extended to compute enthalpy to examine possibility of structural phase transition among these structures. All calculations are performed considering DFT-LCAO method applying the CRYSTAL code [35-38].

The paper is organized as follows: In section 2, we give details on crystal structures, computational methodology and the computational parameters taken in calculations. Results obtained on the structural parameters, pressure-induced structural phase transitions, pressure dependent Compton profiles and AFs are presented and discussed in section 3. In section 4, the results are summarized.

## 2. Crystal structures and computational method

### 2.1 Crystal structures

MgO under ambient conditions crystallizes in the rock-salt i.e. B1 structure belonging to the space group (SG) $Fm\overline{3}m$. The experimental lattice constant of B1 phase is 4.216 Å. This structure is extensively studied compared to others. MgO also crystallizes in the high pressure phase B2 belonging to the SG *Pm3m*. The MgO is presumed to exist in the B3, B4, B8$_1$ and *h*-MgO structures at other pressures. The B3 phase is zincblende with SG $F\overline{4}3m$ while B4 is the wurtzite with SG *P6$_3$mc*. The B8$_1$ phase of MgO has NiAs-type structure while *h*-MgO is buckled boron nitride type structure. The atomic positions of Mg and O in B8$_1$ and *h*-MgO structures are different despite of the same SG *P6$_3$/mmc*.

### 2.2 Computational method

All calculations are performed using first-principles periodic LCAO method implemented in the CRYSTAL code [35-38]. In all electron first-principles self-consistent field (SCF) periodic LCAO procedure, the one electron operator defined as H(**k**) contains kinetic energy operator, Coulomb interaction operator and the exchange operator. The operator transforms to the Fock matrix F(**k**) in reciprocal space. CRYSTAL [35] is based on both Hartree-Fock (HF) and the DFT schemes, allows combination of the two. In the DFT, both exchange and correlation effects are included but approximately in practice. The effectiveness largely depends on the system under investigation. A number of exchange and correlation functionals exist in literature for usage in the first-principles calculations. In the current non-cellular method, each crystalline orbital ψ(**r**,**k**) is a linear combination of Bloch functions φ(**r**,**k**) defined in terms of local functions φ(**r**), normally referred as atomic orbitals. The local functions are expressed as linear combination of certain number of individually normalized Gaussian type functions. For Mg and O, 5-11G [39] and 8-411d1G [40] basis sets were considered. In the present study the Perdew-Burke-Ernzerhof (PBE) [41] exchange and correlation functionals are used. These are based on the generalized gradient approximation (GGA) which essentially accounts for density gradients also that are neglected in the local density approximation. The self-consistent calculations were performed considering Monkhorst and Pack [42] net of 24×24×24 for B1, B3, 21×21×21 for B2, and 18×18×18 points for the B4, B8$_1$ and $h$-MgO structures. The self-consistency was achieved within 11 iterative cycles.

To perform geometry optimization for the wurtzite i.e. B4 structure, the parameters suggested by Schleife *et al* [11] were taken to begin with. The internal parameter *u* was then optimized by performing force calculations (OPTGEOM option) [35,43]. Thus settling the internal parameter *u*, the total energy landscapes were obtained for the (1.45≤(*c/a*)≤1.75) ratio at various volumes. The lowest energy was obtained for *(c/a)*=1.525. At this ratio, the force calculations were performed to refine the internal parameter further. At the refined *u*, the total energy calculations were found stable (tolerance≤0.001 mH) with respect to the *(c/a)* ratio. For the remaining B8$_1$ and $h$-MgO structures the energy landscapes were obtained for (1.6≤(*c/a*)≤1.8) and (1.05≤(*c/a*)≤1.35) respectively. In order to deduce the structural parameters the total energy calculations are coupled with the Murnaghan equation of state for each structure [44].

The Compton profile is an important observable directly related to the EMD distribution in solids. Within validity of the impulse approximation, the spectrum of Compton scattered photons, or Compton profile, is given by [17-19]:

$$J(p_z) = \iint n(\boldsymbol{p}) \mathrm{dp}_x \mathrm{dp}_y, \qquad (1)$$

where n($\boldsymbol{p}$) is the single particle momentum density of the scatterer. Thus Compton profile is a directional property which gives projection of EMD onto a line p$_z$ chosen parallel to the experimental scattering vector (z-axis). For isotropic systems the double integration in Eq. (1) simplifies to:

$$J(q) = 2\pi \int_{|p_z|=q}^{+\infty} \mathrm{pn(p)dp}. \qquad (2)$$

It gives spherically average valence Compton profile. The *n(p)* can be determined from *n($\boldsymbol{p}$)* by the numerical averaging procedure [35,45]. The valence electron momentum density can be determined by sum of the squared moduli of occupied crystalline-orbitals in a momentum representation, or equivalently, by the diagonal element of the six-dimensional Fourier transform of the one electron density matrix from configuration to momentum space using following relations [35-38]:

$$n(\mathbf{p}) = \frac{1}{V_{BZ}} \sum_j^{occ.} \int_{BZ} |\psi_j(\mathbf{k},\mathbf{p})|^2 \, \theta\left[\varepsilon_F - \varepsilon_j(\mathbf{k})\right] d\mathbf{k}, \quad (3)$$

$$= \sum_j \sum_{\mu\nu} e^{-i\mathbf{p}\cdot(\mathbf{s}_\mu - \mathbf{s}_\nu)} a_{\mu j}(\mathbf{p}^0) a^*_{\nu j}(\mathbf{p}^0) \chi_\mu(\mathbf{p}) \chi^*_\nu(\mathbf{p}) \, \theta\left[\varepsilon_F - \varepsilon_j(\mathbf{p}^0)\right], \quad (4)$$

where $\psi_j(\mathbf{k},\mathbf{p})$ is $j^{th}$ occupied crystalline orbital, θ is the step function and the summation extends over all occupied states. In Eq. (4), $a_{\mu j}$ is the expansion coefficient, $p^0$ is the momentum in the BZ, which is related to *p* by a reciprocal lattice vector *K*, $s_\mu$ is the fractional coordinate of the $\chi_\mu$ centre. Fourier transform of the real space atomic orbital $\chi_\mu(r)$ is $\chi_\mu(p)$.

## 3. Results and discussion

### 3.1 Equilibrium structures and the structural phase transitions

The energetics of six phases of MgO were explored by generating total energy curves, E(V), computationally. In case of B8$_1$ and *h*-MgO, *c/a* ratio is optimized while for B4, both the internal coordinate *u* and the *c/a* ratio are optimized [35,43]. The refined internal coordinate for B4 phase obtained after adopting such strategy is *u*=0.3998. The total energy i.e. E(V) curves for the six phases are plotted in Fig. 1. The optimized *c/a* ratio for B4, B8$_1$ and *h*-MgO are 1.525, 1.725 and 1.225 respectively. From Fig. 1 and Table 1 it is visually obvious that equilibrium volume (V$_0$) and total energy at equilibrium decreases as one moves along B4—*h*-MgO—B1. For the B1 structure, computed lattice constant (*a*) is 4.268 Å and bulk modulus (B$_0$) is 160 GPa. These are in very good agreement with the experimental lattice constants *a*=4.216 Å and B$_0$=160 GPa [6,46]. The lattice constants, equilibrium volume, bulk modulus and its pressure derivatives for other structures are listed in Table 1. The current results are in accordance with the previous results. Thus quantitative reliability in structural parameters aptly ensures stability and accuracy of the basis sets considered in this study.

In order to examine possibility of structural phase transitions and compute transition pressure ($P_t$), the enthalpy (H=E+pV) calculations were performed. In Fig. 2, enthalpy (H) curves for all six polymorphs are plotted in two panels. The left panel shows enthalpy within the pressure range -25—75 GPa while the right panel shows enthalpy within the range of 300—450 GPa. In the 75—300 GPa range no pressure dependent transitions are seen. Both panels in Fig. 2, depict that B1→B2, B3→B2, B4→B2, B8$_1$→B2, $h$-MgO→B2, $h$-MgO→B8$_1$ and B4→B8$_1$ transitions are permissible on applying pressure. Most of the transitions are found to occur below 75 GPa. The B4→B8$_1$, $h$-MgO→B8$_1$, B3→B2, B4→B2 and $h$-MgO→B2 transitions take place at 2, 9, 37, 42 and 64 GPa respectively. Possibilities of some of these transitions are predicted by other workers without reporting transition pressure [11]. The high pressure transitions B1→B2 and B8$_1$→B2 are found to occur at 410 and 340 GPa respectively. Notably, only B1→B2 transition is studied experimentally [1-3]. Earlier it was found that B1 is a stable up to 110 GPa [1,6]. Later experiment showed that B1 is stable up to 227 GPa [2,3]. Theoretically the B1→B2 transition is studied by many workers [4-7, 10-14]. The transition pressure predicted by current calculation is well in accordance with earlier predictions. In the current study six more transitions are found which take place at negative pressure (-20—0 GPa). It may be noted that negative pressure cannot be realized in experiments and therefore have limited practical applications.

### 3.2 Pressure dependent Compton profiles:

The total, core and valence isotropic Compton profiles for the B2 and B4 structures computed at 0, 25, 42, and 75 GPa are plotted in Figs. 3(a-d). One observes that core electrons profiles are independent of the pressure applied. The change in the total Compton profile is entirely governed by change in the valence electrons profile. These findings are also observed in pressure dependent profiles computed for other polymorphs of MgO. Therefore, in Fig. 4 the pressure dependent valence electrons profiles, typically, for the B1 and $h$-MgO structures are plotted. Valence Compton profiles of remaining structures show similar trends. Figures 3 and 4 reveal that the J(0) decreases gradually as pressure increases. Change in J(0) is maximum when the pressure is low (say 0-25 GPa in this case). This is well expected because in this region relative change in volume is more compared to the high pressure region for an equal increment in pressure.

Moreover, increase in pressure broadens the Compton profiles. This broadening is also well anticipated. On the basis of band structure calculations of B4, $h$-MgO, and B1 structures, Limpijumnong and Lambrecht [8] have pointed out that decrease in volume results in an increase in the valence band width accompanied by a downward shift of the centre of gravity of the major valence bands. These two effects collectively broaden the Compton profiles with decrease in volume or increase in pressure. A remarkable feature common in the Compton profiles is the notion that the $\left.\frac{dJ(q)}{dP}\right|_{q=0.7} = 0$, where P is the pressure in GPa. Below q=0.7 a.u., the Compton profiles decrease with pressure and increase thereafter. This is clearly visible in the highlighted region of the curves given in the inset of Fig. 4. Notably, the position of this point is independent of the structures of MgO. It may probably be species dependent. Similar studies on other materials are necessary to see its origin.

Further salient fine features on the EMD and the pressure dependence thereon can be gained by computing the first and second derivatives of the isothermal Compton profiles. The first derivative of Compton profile is related to the 3-D EMD as [19,47,48]:

$$n(q) = \frac{-1}{2\pi q} \frac{dJ(q)}{dq}, \qquad (5)$$

where J(q) is the average Compton profile. Likewise the second derivative in Compton profile is related to n(0) as [19,47]:

$$n(0) = \frac{-1}{2\pi} \left.\frac{d^2 J(q)}{dq^2}\right|_{q=0}. \qquad (6)$$

The first and second derivatives of Compton profiles of B1 and $h$-MgO structures at various pressures (-25 to 50 GPa) are plotted in Figs. 5(a) and 5(b) respectively. It is known that both first and second derivatives show rapid change at the Fermi momentum ($q_F$). For both structures of MgO, the momentum of the principal maxima where rapid changes are visible shift towards higher momentum side with pressure pointing that Fermi momentum increases with pressure. This is in conformity with the experimental evidence derived from the Compton profile measurements on free electron gas of sodium at high pressure. Because of the relatively high electron density in Na, shift in the $q_F$ occurs at lower pressure in contrast to the ionic solid MgO. Moreover shift in $q_F$ with pressure is less in MgO than Na. Interestingly, only the second derivatives in Compton profiles, drawn in Fig. 5 (b), show specific signatures at q=0.7 a.u. At -25 and 0 GPa it shows regular behavior and a deep starts to appear thereafter in the second derivatives for the B1 structure. This deep is shallower in $h$-MgO and becomes pronounced at higher pressure.

Among several polymorphs of MgO, the B1→B2 transition is experimentally reported to occur at pressure greater than 227 GPa. Most of the theoretical calculations observed this transition within the 400-600 GPa range of pressure. Therefore we compare the valence Compton profiles of the B1 and B2 structures in Fig. 6. The given Compton profiles are computed below transition pressure i.e. 100 GPa, at transition pressure i.e. 410 Gpa, and after transition pressure i.e. 475 GPa. It is obvious that the pressure causes broadening of the profiles. This is well along the findings of Chang and Cohen [4] who observed broadening in the valence bands widths of the B1 and B2 polymorphs with decrease in volume in the band structure calculations. One can see some change in Compton profiles of B1 and B2 structures between $0.0 \leq q \leq 1.0$. These changes are different from the changes observed in Fig. 3. This may probably due to the pressure induced structural phase transition from B1 to B2. For further analysis, the first derivatives of the valence Compton profiles of the two structures at 100, 410 and 475 GPa are plotted in Fig.7 (a). At transition pressure the principal maxima reduces in amplitude and another maxima, low in amplitude, appears around 0.3 a.u. in case of B1. The low amplitude maxima is absent below and above the transition pressures. On the other hand, a maxima appears at transition pressure and remains at 475 GPa for the B2 phase. These features are more clearly visible in the second derivatives plotted in Fig. 7(b). Both the first and second derivatives below and above transition pressure clearly hints at reorganization of valence electron momentum density during structural phase transition. Finer details can be gained by observing the changes in anisotropies in the directional Compton profiles. It is hoped that the present study will open avenues to study pressure dependent Compton profiles theoretically and experimentally.

### 3.3 Pressure dependent autocorrelation functions:

The Fourier transform of the Compton profile is related to the autocorrelation function (AF) or the B-function by the following relation [17-19,36]:

$$B(r) = \int_{-\infty}^{\infty} J(q) \exp(iqr)\,dq. \tag{7}$$

The AF function has the property that B(0)=N, where N is the number electrons contributing to the Compton profile. The AF introduced by Pattison and William [49] has the advantage that one deals back in the real space and can focus on the region dominated by the valence electrons. The position of deep and zero crossings are related to the bonding distances and nearest neighbour interaction in the solids [36,48,49]. The isotropic AFs of the B1 structure at various pressures are plotted in Fig. 8. Two features are noteworthy. Firstly, under ambient conditions, B(r) = 0 at r= 3.55 a.u., the local minimum occurs at the 4.6 a.u. and a shoulder is visible at 7.2 a.u. Beyond this, all curves show nearly identical behaviour. It shows that the charge transfer occurs mostly within the $3.5 \leq r \leq 7.5$ range. Increment in pressure shifts these positions towards the lower r side. The high pressure i.e. compression brings atoms closer thereby reducing the lattice constant. It reduces the nearest neighbour distances and the bond length. This is probably the reason for shifting towards the lower r side. Secondly, a typical AF has a minimum value at a characteristic r. In Fig. 8, the minimum value of AF goes further down with increment in pressure. The depth and height in the AF after first zero crossing are attributed to the strength of interaction among electrons of nearest neighbours forming bonds [49]. An increase in the depth and height with pressure indicates stronger bonding. Also the value of AF at the shoulder increases with increment in pressure. The O 2p states have been attributed a major part in the bonding of MgO in B1 structure on the basis of AF by a number of workers [31-33]. The O-2s states play relatively a minor role. The interaction between bonding states becomes more pronounced in terms of charge transfer or overlap leading to a larger change in the autocorrelation functions as pressure increases. These features can be better analysed in terms of directional AFs.

## 4. Conclusions

The first-principles method at the level of DFT is applied to study structural phase transitions in six polymorphs of MgO. The calculations are performed to see effect of pressure on the Compton profiles and the autocorrelation functions. The enthalpy calculations predict that B4→B8$_1$, h-MgO→B8$_1$, B3→B2, B4→B2 and h-MgO→B2 transitions take place at 2, 9, 37, 42 and 64 GPa respectively. The high pressure transitions B8$_1$→B2 and B1→B2 are found to occur at 340 and 410 GPa respectively. It is observed that the valence Compton profiles are sensitive to the applied pressure. The core profiles are almost independent of the pressure in all MgO polymorphs. The valence electrons profile becomes broad with increment in pressure. With the increase in pressure the principal maxima in the second derivative of Compton profiles shifts towards high momentum side in all structures. The first and second derivatives before and after the transition pressure suggest reorganization of momentum density in the B1→B2 structural phase transition. Features of the autocorrelation functions shift towards lower r side with increment in pressure. The possibility of theoretical Compton profile calculations presented in this work may invite further pressure dependent studies. It will also enable to establish Compton scattering as more versatile probe to study properties of materials under different conditions like magnetic field and pressure.

## Acknowledgements

UP is grateful to the Council of Scientific and Industrial Research, New Delhi for awarding Senior Research Fellowship. Financial support provided by the UGC, New Delhi through grant No. SR/33-37/2007 to BKS is gratefully acknowledged. BB is supported by the U.S. Department of Energy under Contract Nos. DE-FG02-07ER46352 and DE-SC0007091.

**Figure captions**

**Fig.1** Volume dependent total energy curves for six MgO polymorphs.

**Fig. 2** The computed enthalpies for six MgO polymorphs.

**Fig. 3** Total, core and valence electrons Compton profiles of B2 and B4 phases of MgO under (a) ambient conditions (b) 25 GPa, (c) 42 GPa, and (d) 75 GPa.

**Fig. 4** Pressure dependent valence Compton profiles of B1 and *h*-MgO phases. The inset shows the highlighted region around q=0.70 a.u.

**Fig. 5** Pressure dependence of the (a) first and (b) second derivatives of valance profiles of B1 and *h*-MgO phases.

**Fig. 6** Valence electrons Compton profiles of B1 and B2 phases at higher pressure.

**Fig. 7** Pressure dependence of the (a) first and (b) second derivatives of valance profiles of B1 and B2 phases at higher pressure.

**Fig. 8** Pressure dependent autocorrelation functions of B1 phase of MgO.

**Table 1.** Lattice parameters, Volume ($V_0$), bulk modulus ($B_0$) and its pressure derivative ($B'$) for various phases of MgO.

| Phase | | | $V_o$ | $a$ (Å) | $c$ (Å) | $B_0$ (GPA) | $B'$ |
|---|---|---|---|---|---|---|---|
| B1 | Expt. | | | 4.216 [a] | | 160.3 [a] | |
| | | | | | | 177.0 [b] | 4.0 [b] |
| | Theory | This Work | 19.442 | 4.268 | | 160.4 | 3.43 |
| | | PP-VASP [c] | -- | 4.254 | | 148.6 | 4.30 |
| | | LCAO-GGA [d] | 19.156 | 4.247 | | 169.1 | 3.28 |
| | | LCAO-LDA [d] | 18.034 | 4.163 | | 185.9 | 3.40 |
| | | DFPT [e] | 18.810 | 4.222 | | 159.0 | 4.30 |
| | | PAW-GGA [f] | 19.012 | 4.237 | | 154.2 | 4.14 |
| | | HF-LCAO [g] | 18.400 | 4.200 | | 182.0 | 3.92 |
| | | FP-LMTO [h] | 17.800 | 4.145 | | 178.0 | |
| | | DFT-GGA [i] | 18.760 | | | 161.5 | 4.00 |
| B2 | Theory | This work | 18.992 | 2.668 | | 156.3 | 3.21 |
| | | PP-VASP [c] | -- | 2.661 | | 140.3 | 4.10 |
| | | LCAO-GGA [d] | 18.740 | 2.656 | | 152.6 | 3.39 |
| | | LCAO-LDA [d] | 17.573 | 2.599 | | 169.8 | 3.54 |
| | | HF-LCAO [g] | 17.600 | 2.600 | | 181.0 | 4.00 |
| | | DFT-GGA [i] | 18.200 | | | 145.5 | 4.05 |
| B3 | Theory | This work | 24.203 | 4.592 | | 148.1 | 2.99 |
| | | PP-VASP [c] | | | | | |
| B4 | Theory | This work | 24.103 | 3.317 | 5.059 | 148.4 | 2.95 |
| | | PP-VASP [c] | | 3.322 | 5.136 | 116.9 | 2.70 |
| | | FP-LMTO [h] | 22.500 | 3.169 | 5.175 | 137.0 | |
| $h$-MgO | Theory | This work | 22.745 | 3.500 | 4.288 | 142.0 | 3.15 |
| | | PP-VASP [c] | | 3.523 | 4.236 | 124.8 | 4.30 |
| | | FP-LMTO [h] | 20.900 | 3.426 | 4.112 | 148.0 | |
| $B8_1$ | Theory | This work | 19.818 | 2.983 | 5.150 | 151.3 | 3.38 |

[a] Ref. 46
[b] Ref. 2
[c] Ref. 11
[d] Ref. 7
[e] Ref. 50
[f] Ref. 51
[g] Ref. 6
[h] Ref. 8
[i] Ref. 3

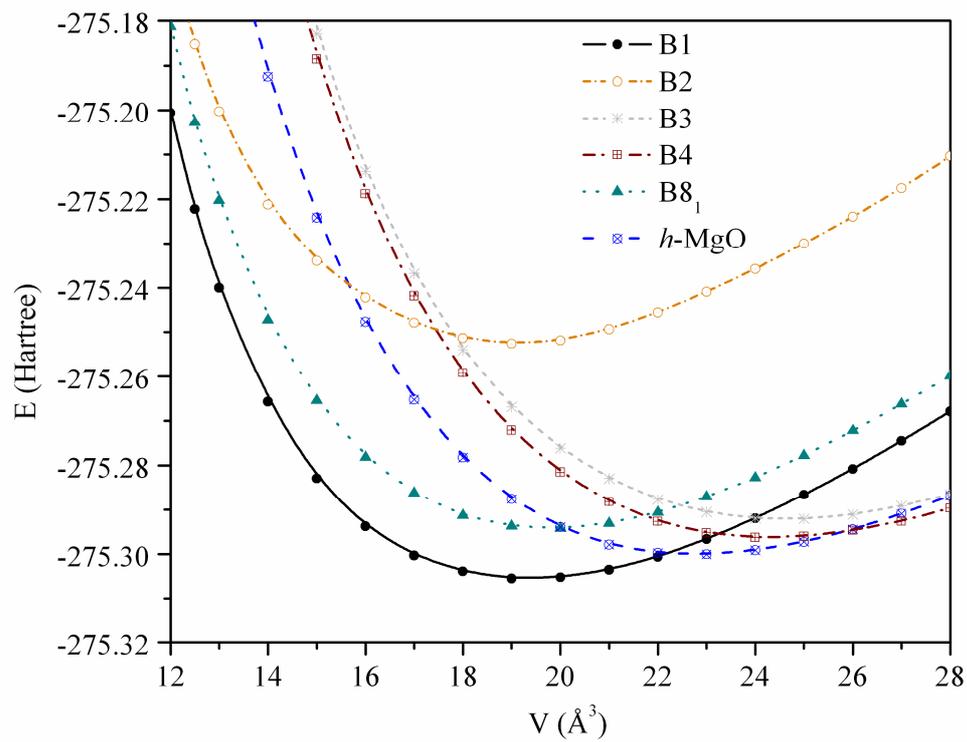

**Figure** 1.

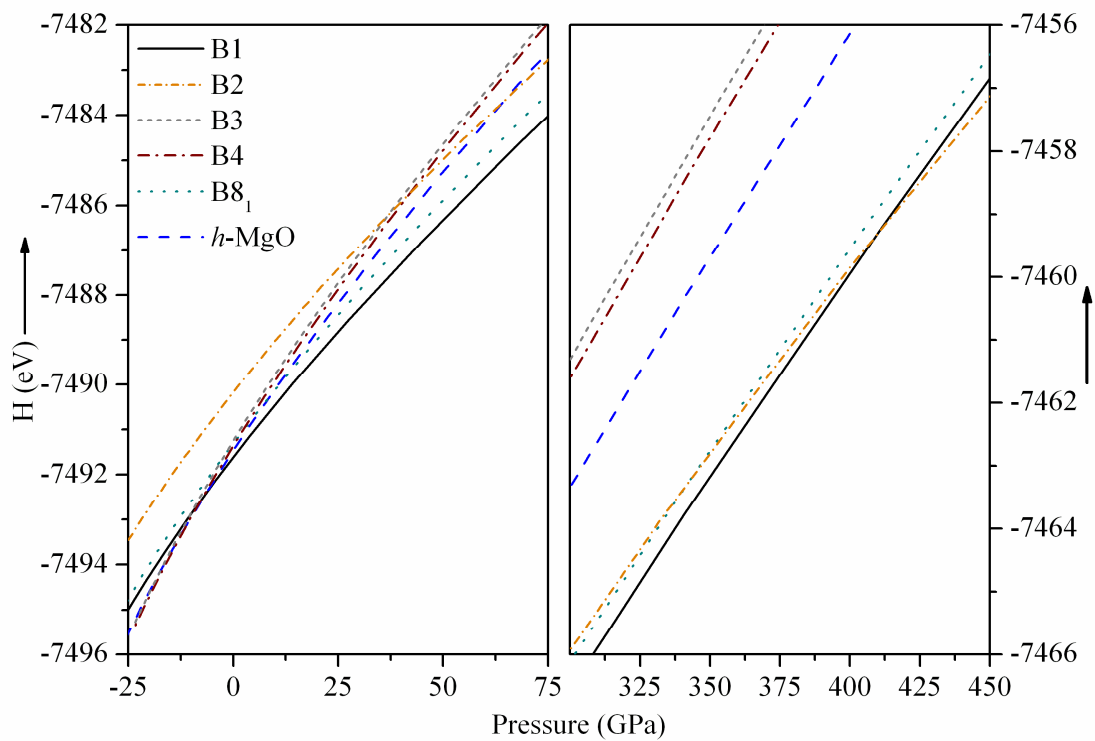

**Figure** 2.

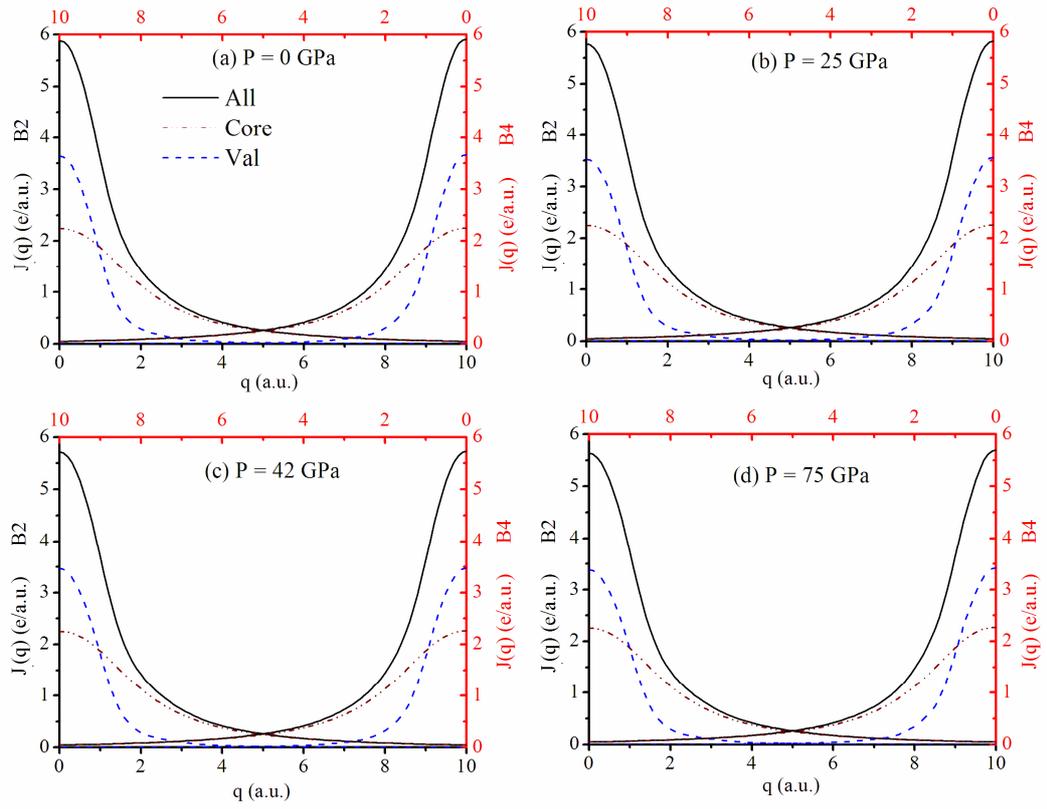

**Figure** 3.

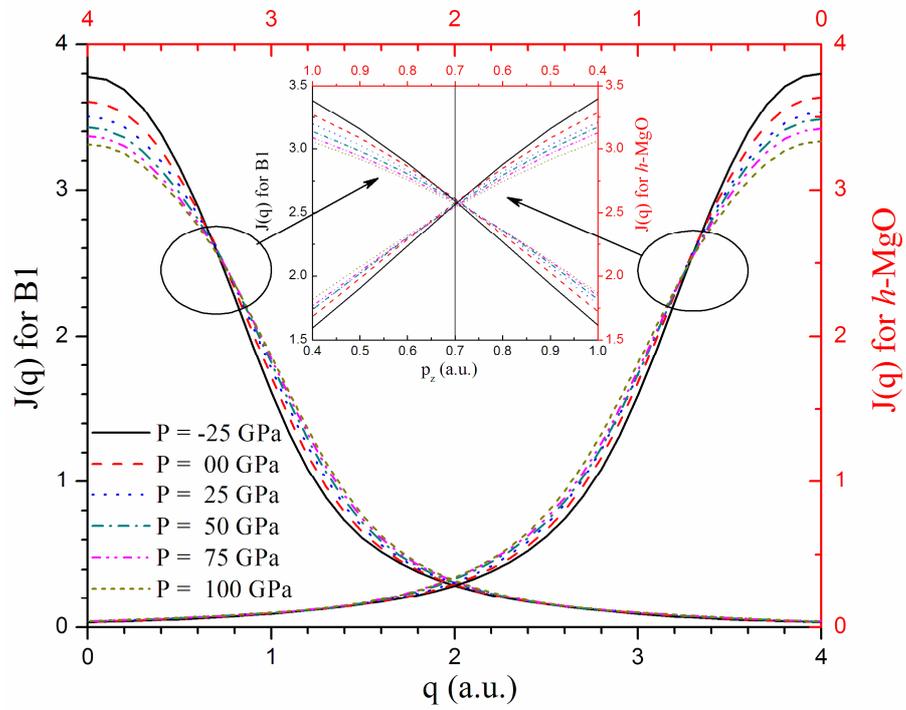

**Figure** 4.

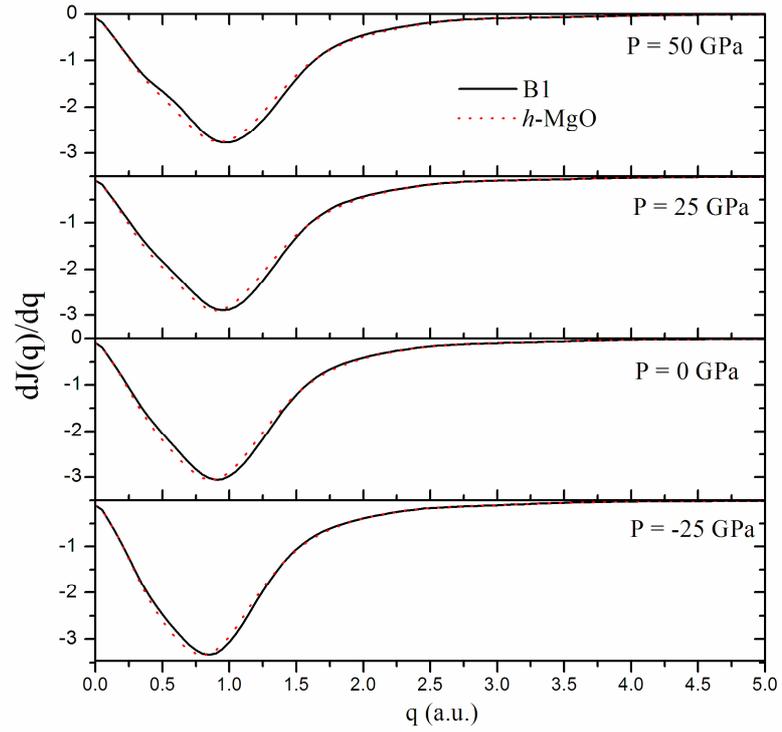

**Figure** 5(a).

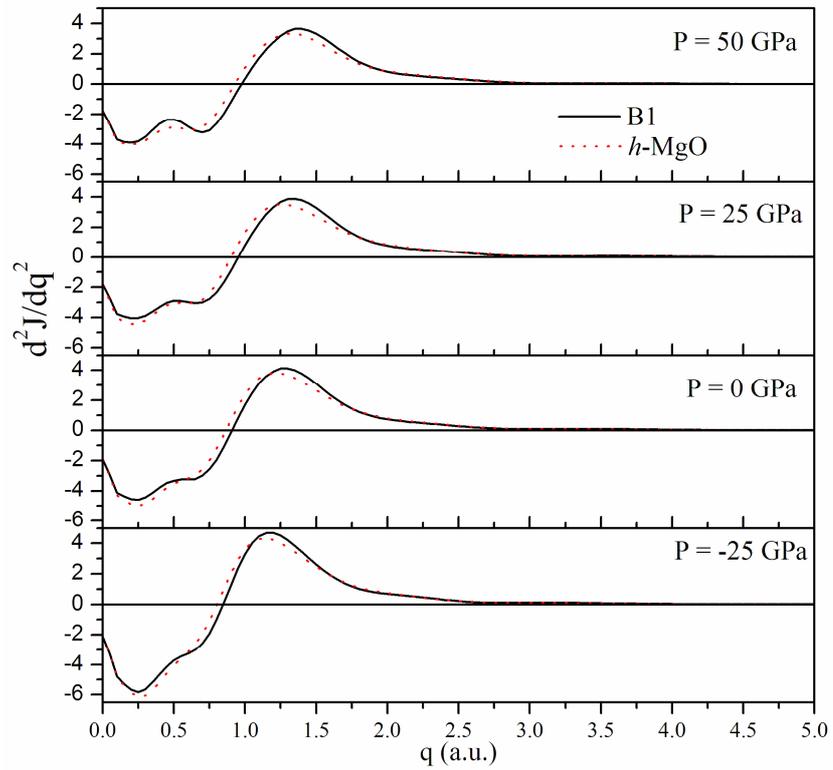

**Figure** 5(b).

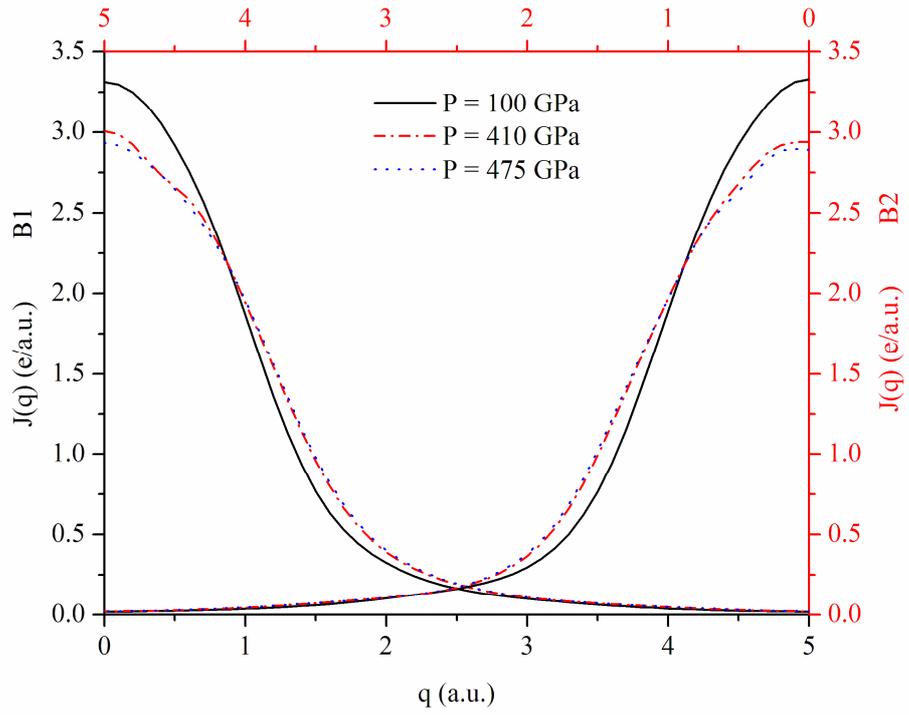

**Figure** 6.

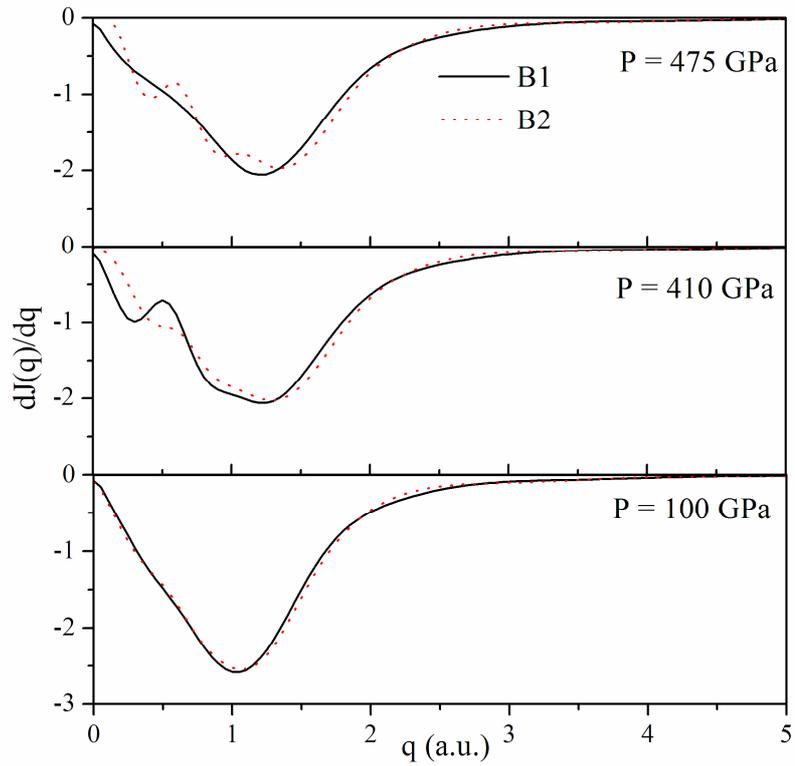

**Figure** 7a.

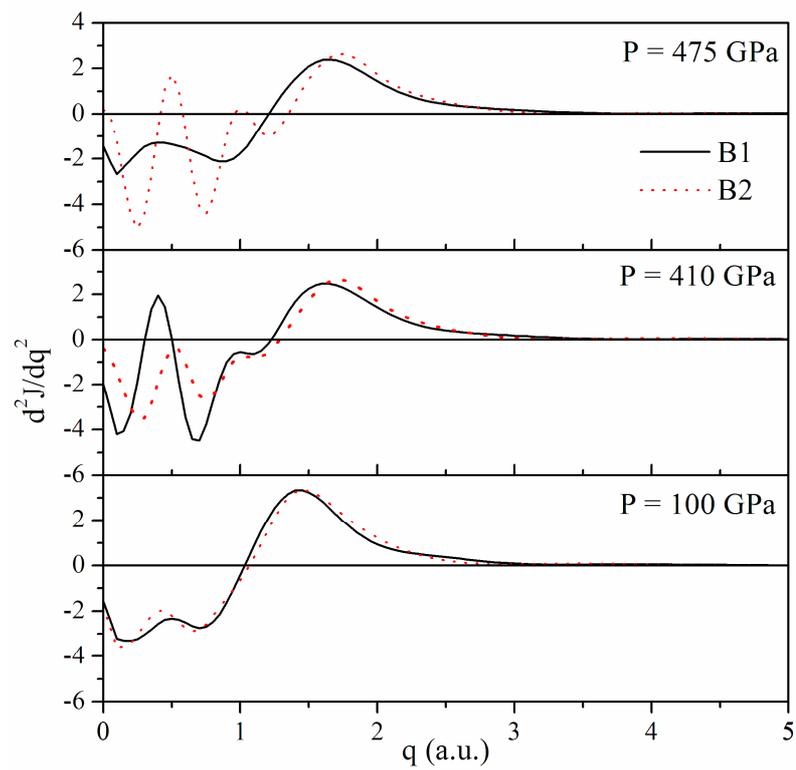

**Figure** 7b

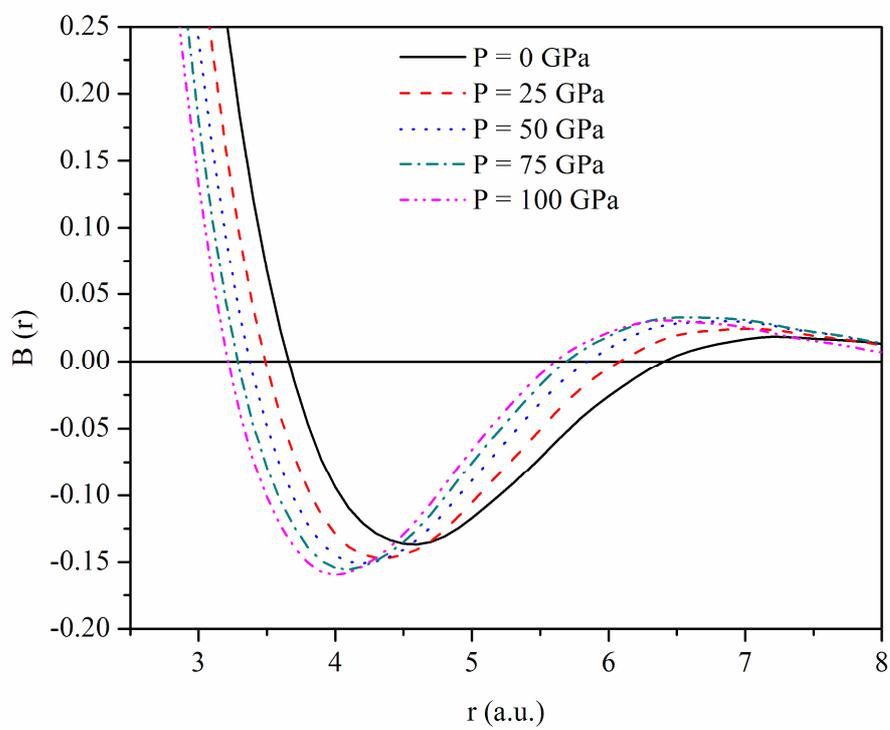

**Figure** 8